# Mitigation of relative intensity noise of Quantum Dash mode-locked lasers for PAM4 based optical interconnects using encoding techniques

**VIDAK VUJICIC,[1]\* ARAVIND P. ANTHUR,[1] ARSALAN SALJOGHEI,[1] VIVEK PANAPAKKAM,[2] RUI ZHOU,[3] QUENTIN GAIMARD,[2] KAMEL MERGHEM,[2] FRANCOIS LELARGE,[4] ABDERRAHIM RAMDANE,[2] AND LIAM P. BARRY[1]**

[1]*Department of Electronic Engineering, Dublin City University, Dublin 9, Ireland*
[2]*Centre de Nanosciences et Nanotechnologies (C2N), CNRS, Marcoussis 91460, France*
[3]*Huawei Technologies, Shenzhen City, Guangdong, China*
[4]*III–V Lab, Joint Laboratory of Nokia Bell Labs and Thales Research and Technology, CEA-LETI, Marcoussis 91460, France*
*\*vidak.vujicic@dcu.ie*

**Abstract:** Quantum Dash (Q-Dash) Passively Mode-Locked Lasers (PMLLs) exhibit significant low frequency Relative Intensity Noise (RIN), due to the high Mode Partition Noise (MPN), which prevents the implementation of multilevel amplitude modulation formats such as PAM4. The authors demonstrate low frequency RIN mitigation by employing 8B/10B and Manchester encoding with PAM4 modulation format. These encoding techniques reduce the overlap between the modulation spectral content and the low-frequency RIN of the Q-Dash devices, at the expense of increased overhead. The RIN of the 33.6 GHz free spectral range Q-Dash PMLL was characterized, and the results obtained show very high levels of RIN from DC to 4 GHz, but low levels for higher frequencies. The performance improvement for 28 GBaud 8B/10B and Manchester encoded PAM4 signal has been demonstrated compared to the case when no encoding is used. Finally, the effect of RIN on the system performance was demonstrated by comparing the Bit Error Rate (BER) performance of the PAM4 signaling obtained with an External Cavity Laser (ECL) to those obtained with Q-Dash PMLL.

## 1. Introduction

Internet traffic shows a relatively steady growth rate over the last 15 years, in the range of 20-50% per year [1,2], mainly driven by various cloud services processed within data centers in the last decade. As system capacity scales to meet the growth of network traffic, it is important for the speed of interface rates to the network line cards to increase in order to restrict the upsurge in complexity of the network [1], which is mainly important for data centers. To achieve that, standardization of 50G, 100G, 200G and 400G Ethernet interfaces based on 56 Gb/s serial data rates is currently in progress and is expected to be finished by 2020 [3]. Although a number of modulation formats have been considered as potential candidates for future short reach communications [4-8], 4-level Pulse Amplitude Modulation format (PAM4) has gained significant attention from the Ethernet community mainly due to similarities with NRZ modulation format [9-11].

Besides requirements for higher data rates per interface, there is a constant demand for reduction of size and power consumption of data center interconnects. A fully integrated optical interconnect transceiver incorporating active and passive components on a single chip is an attractive solution to meet the above requirements [12,13]. Nowadays, short-range optical interconnects rely on arrays of Vertical Cavity Surface Emitting Lasers (VCSELs), Distributed Feedback (DFB) lasers and Electro-absorption Modulated Lasers (EML) as optical sources, each producing one of the several optical carriers with high power and low Relative Intensity Noise (RIN) [9,14,15]. Nevertheless, several limitations due to the practical implementation of parallel laser arrays occur, such as "smile error", requirement for emission frequency stabilizers at each carrier, high insertion losses of on-chip multiplexers and potential high cost of implementation of large number of lasers into a single module [4,13]. An attractive alternative in developing Wavelength Division Multiplexed (WDM) data center interconnect solutions is based on a single Quantum Dash (Q-Dash) Passively Mode-Locked Laser (PMLL) source producing a frequency comb with a fixed carrier spacing [4,13,16,17]. Although this solution is compact and conceptually simple, the high level of low frequency RIN due to the effect of Mode Partition Noise (MPN), which is characteristic for Q-Dash MLLs, limits the number of modulation formats that can provide sufficiently good performance [4,18-20].

Powerful techniques for mitigation of MPN which have been presented previously are based on an injection-locking mechanism [21] and on amplifying the filtered lines of MLL with a saturated Semiconductor Optical Amplifier (SOA) in order to filter the low frequency part of the RIN spectrum [19]. However, those solutions are not practical as they require additional optical components (optical sources and amplifiers) operating under strict conditions, which increase total complexity, power consumption and size. Furthermore, they are not compatible with serial modulator architectures based on ring resonator modulators that have been recently proposed [13]. Also, RIN reduction in analogue links using balanced detection has been reported [22]. However, balanced detection is not compatible with the proposed integrated optical interconnects which use single-ended detection [12,13].

In this paper we investigate low frequency RIN mitigation by employing 8B/10B and

Manchester encoding with PAM4 modulation format. These encoding techniques reduce the overlap between the modulation spectral content and the low-frequency RIN, at the expense of increased overhead. Encoding of PAM4 signal with 8B/10B and Manchester techniques enables performance improvement compared to the case when no encoding is employed. For the case of uncoded PAM4 signaling, the performance of filtered channels from the Q-Dash PMLL fluctuated around 7% Forward Error Correction (FEC) limit. However, when the encoding was used, the majority of the channels exhibited performance better than FEC limit. Manchester encoding provides better performance than 8B/10B, but has significantly reduced net spectral efficiency for a given data rate due to the encoding overhead. Detailed RIN analysis of the employed Q-Dash PMLL, grown on an S-doped (001) InP substrate with an active region composed of six layers of InAs Q-Dashes separated by InGaAsP barriers, is also presented.

## 2. Principles of 8B/10B and Manchester encoding for PAM4

Various encoding techniques can be used to alter the properties of a signal's spectral content by rearranging its binary sequence. Line codes that suppress the DC component of a signal can be beneficial in a several aspects as they aid clock recovery at the receiver [23], provide spectral notches at desired frequencies for integration of secondary signals [24] and overcome the issue that many high bandwidth components have minimum frequency cut-offs in the MHz range [25]. The suppression of low frequency spectral content, i.e. generation of a spectral notch from DC to a certain frequency, can be used for low frequency noise mitigation. In this paper we investigate low frequency RIN mitigation using 8B/10B and Manchester encoding schemes. 8B/10B encoders have been applied previously on 4-level signals in order to avoid the penalty caused by the loss of the low-frequency components. Nevertheless, the encoded 4-level electrical signal was used for the generation of QPSK signal at the transmitter [25], and was not detected as PAM4 signal using direct detection. This is the first application of 8B/10B encoding to PAM4 based optical communication system to the best of our knowledge. Moreover, to the best of our knowledge Manchester encoded PAM4 signaling has not been demonstrated before in optical communication systems.

Figure 1 illustrates the principle behind generation of 8B/10B and Manchester encoded PAM4 signals. In the case of 8B/10B encoded PAM4 signal, two binary sequences are separately encoded using the 8B/10B encoder as shown in Fig. 1(a). The encoded data sequences are then combined and the resulting data is mapped to the PAM4 modulation format. Due to the coding overhead, net spectral efficiency of 8B/10B encoded PAM4 signal is 20% lower for a given data rate compared to uncoded PAM4. An eye diagram of the measured 28 GBaud 8B/10B PAM4 signal is also shown in Fig. 1(a). For the purpose of eye diagram measurements, an External Cavity Laser (ECL) was externally modulated with the 8B/10B PAM4 signal and then detected with a 20 GHz PIN photoreceiver before being captured with a Tektronix real-time oscilloscope (RTO).

In the case of Manchester encoded PAM4 signal, a binary sequence is first mapped to PAM4 constellation, as illustrated in Fig. 1(b). Afterwards, transitions between the signal levels were generated to reflect the Manchester encoding, as shown in Fig. 1(b), i.e. low-to-high transitions -3 to 3 and -1 to 1, and high-to-low transitions 3 to -3 and 1 to -1. The PAM4 data and corresponding transitions were used to modulate the optical carrier. Therefore, due to the transitions overhead, net spectral efficiency of Manchester encoded PAM4 signal is halved for a given data rate compared to uncoded PAM4. An eye diagram of measured 28 GBaud Manchester PAM4 signal is shown in Fig. 1(b) as well. Similar to the case of 8B/10B encoding, an ECL was externally modulated with the Manchester PAM4 signal and then detected with a 20 GHz PIN photoreceiver before being captured with a Tektronix RTO.

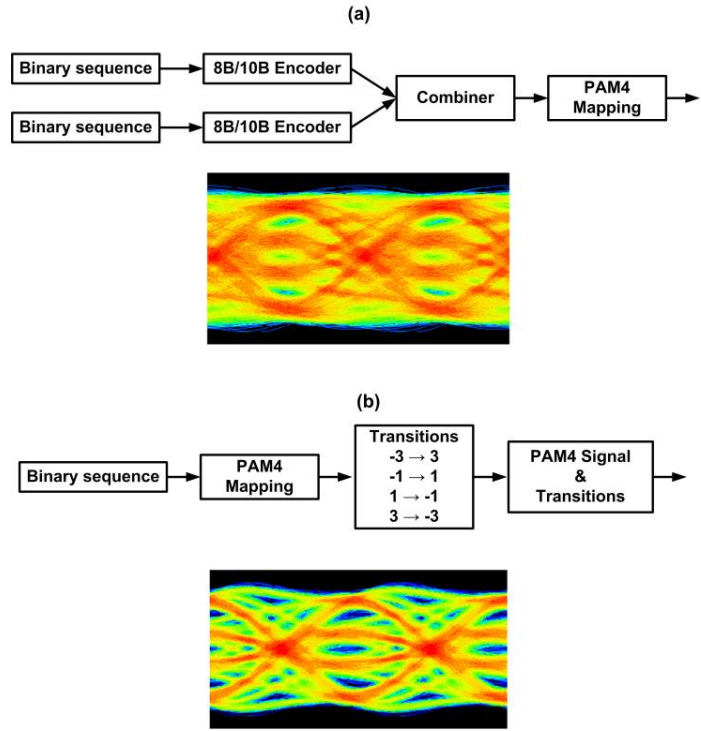

Fig. 1. (a) Schematic of 8B/10B encoder: Two binary sequences are separately encoded before being combined and mapped to PAM4. (b) Manchester encoder: A binary sequence is mapped to PAM4 constellation before symbol transitions are generated and combined with PAM4 data.

Figure 2 shows the measured spectral content of the 28 GBaud PAM4 signal without any encoding and with 8B/10B and Manchester encoding applied. The spectral content of the 28 GBaud PAM4 signal was shifted away from the low frequency region when encoding was applied. The signal spectra were measured using an Electrical Spectral Analyzer (ESA) directly from an Arbitrary Waveform Generator (AWG) which has 25 GHz analog bandwidth. 8B/10B encoding generates a spectral notch of a few hundred MHz, whilst Manchester encoding provides a spectral notch with a bandwidth of a few GHz. The width of the spectral notch is solely dependent on the encoding properties of the particular encoder employed [26]. The width of the spectral notch is fixed for 8B/10B and Manchester encoders. The property of the investigated encoding techniques is that the encoders which introduce larger overhead also provide larger bandwidth of the spectral notch. Therefore, based on the level of low frequency RIN, we need to choose the encoding scheme which ensures the transmission performance criteria is met, with the lowest overhead.

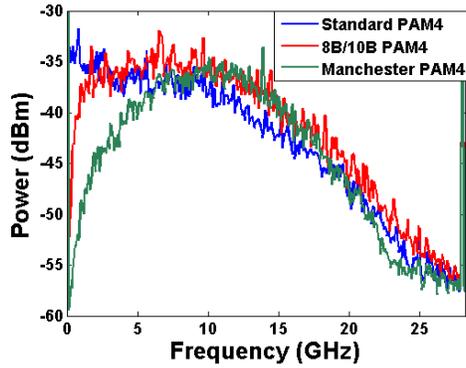

Fig. 2. RF spectra of standard 28 GBaud PAM4 signal (blue), 8B/10B PAM4 signal (red) and Manchester PAM4 signal (green) captured with an Electrical Spectral Analyzer (ESA) directly from an Arbitrary Waveform Generator (AWG). 8B/10B encoding generates a spectral notch of a few hundred MHz, whilst Manchester encoding provides a spectral notch with a bandwidth of a few GHz.

## 3. Q-Dash device and characterization

The device used in the experiment is a single-section Q-Dash based laser. The device was grown by a Gas Source Molecular Beam Epitaxy (GSMBE) on an S-doped (001) InP substrate. The active region consists of six layers of InAs Q-Dashes separated by InGaAsP barriers. This design provides a low optical confinement factor, resulting in a reduced impact of spontaneous emission on the intensity and phase noise of the laser longitudinal modes [27]. These buried-ridge stripe waveguide devices were processed with a ridge width of 1.5 μm. The as-cleaved laser used here has a total length of 1250 μm, corresponding to repetition frequency of 33.6 GHz. The laser was temperature controlled throughout the measurements at 23°C using a thermoelectric cooler (TEC) module, which had an accuracy of ±0.1°C. When biased with a single current source at 350 mA the laser exhibited the characteristic square-shaped emission spectra [4] with a spectral bandwidth in excess of 1.5 THz, as shown in Fig. 3(a). The applied bias currents provided an average power in fiber of 9 dBm. The measured RF linewidth of the 33.6 GHz beat signal was 90 kHz, indicating strong passive mode-locking [28]. The detected RF spectrum is shown in Fig. 3(b).

Characterization of the intensity noise is critically important as semiconductor mode-locked lasers can exhibit significant MPN [4], which could impair system performance [4,29]. The RIN of the Q-Dash device was determined for selected individual spectral modes and also for the entire emission spectrum. Measured RIN spectra over 30 GHz bandwidth are shown in Fig. 3(c), and RIN spectra measured over 5 GHz are given in Fig. 3(d). In all cases the individual spectral modes possess high intensity noise at frequencies below 4 GHz, but low intensity noise at higher frequencies. However, the measured RIN of the entire emission spectrum is very low and comparable with RIN of the ECL used (Agilent N7711A). This disparity indicates the presence of high MPN in this device. Furthermore, Optical Carrier-to-Noise Ratio (OCNR) values of the comb lines used for data transmission experiment has been measured to be in excess of 45 dB, using an Optical Spectrum Analyzer (OSA) with resolution bandwidth (RBW) of 0.02 nm.

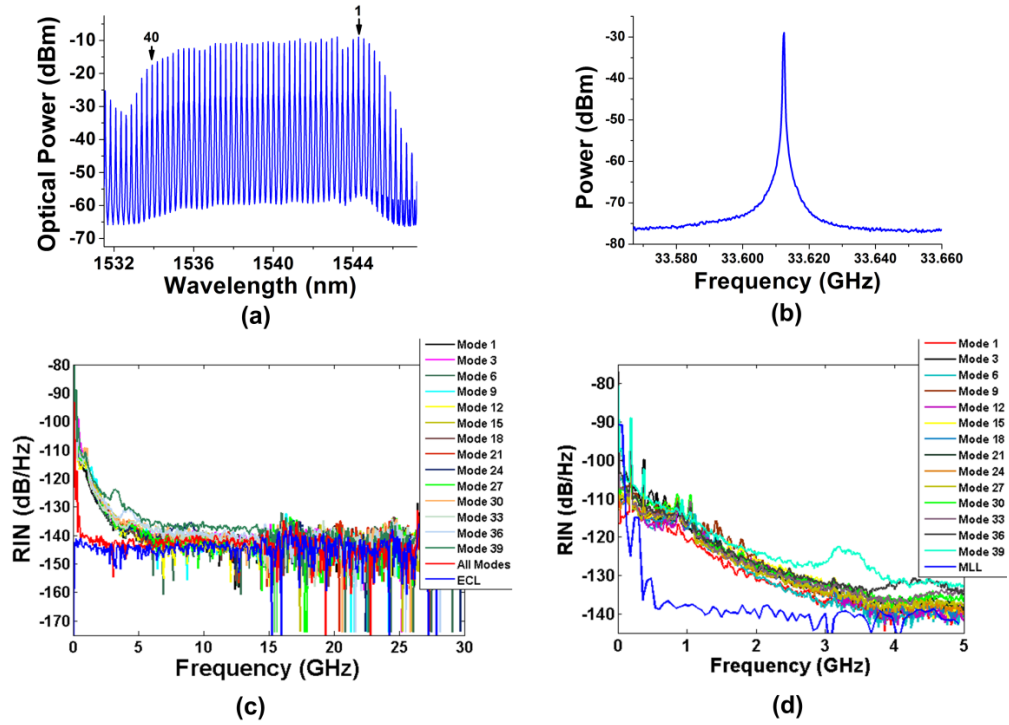

Fig. 3. (a) Optical spectrum of 33.6 GHz Q-Dash PMLL (Resolution bandwidth (RBW) was 0.02nm). (b) Detected RF linewidth. (c) RIN for selected filtered modes over 30 GHz bandwidth, all modes and ECL and (d) RIN for selected filtered modes over 5 GHz bandwidth and all modes.

## 4. Experimental setup

The experimental setup for the PAM4 system is shown in Fig. 4. The Q-Dash PMLL was used to generate ~ 40 optical carries, which were individually filtered using an optical bandpass filter (OBPF) with a sharp roll-off and a 3-dB bandwidth of around 15 GHz, enabling efficient suppression of adjacent carriers. The output of the OBPF is amplified with an Erbium doped fiber amplifier (EDFA) and the Amplified Spontaneous Emission (ASE) noise from the EDFA is filtered using a 2 nm tunable optical filter. The filtered comb line was modulated by a single-drive Mach-Zehnder modulator (SD-MZM), which was biased at the quadrature point and then used to modulate the filtered comb line with an amplified 28 GBaud PAM4 signal waveform generated from an arbitrary waveform generator (Keysight M8195A) operating at 56 GSa/s. PAM4 signal waveforms were generated using $2^{16}$ bits long Pseudo Random Binary Sequence (PRBS) and were pre-distorted to compensate for nonlinearities of the data driver and SD-MZM. The optical spectrum of a filtered modulated comb tone is shown in inset in Fig. 4.

At the receiver side, the filtered channel was detected using a 20 GHz receiver that consists of a PIN photodetector and an integrated trans-impedance amplifier (TIA), after being transmitted over 3 km of standard single mode fiber (SSMF). The received signal was captured with a 33 GHz Tektronix DPO77002SX RTO operating at 100 GSa/s. Digital processing of the received signal (resampling, normalization, adaptive equalization, symbol synchronization, decoding), and Bit Error Rate (BER) calculations, were performed offline using Matlab. The adaptive equalizer was a 13 tap Finite Impulse Response (FIR) filter, and the tap weights were updated using Decision-Directed Least-Mean Square (DD-LMS) algorithm [30]. The symbol synchronization was performed with the aid of 32 PAM4 symbols long training sequence. In the case of 8B/10B PAM4 signal, an 8B/10B decoder was applied after PAM4 decoder, whilst in the case of Manchester PAM4 signal decoders were applied prior to PAM4 decoder.

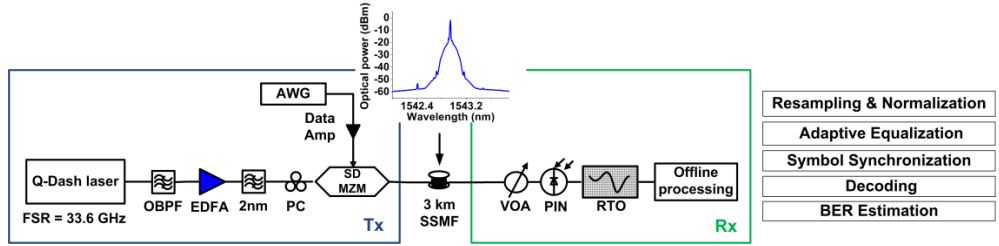

Fig. 4. Schematic of the experimental setup and DSP flow at the receiver. Inset shows optical spectrum of a filtered modulated comb tone. Optical Band-Pass Filter (OBPF); Polarization Controller (PC); Arbitrary Waveform Generator (AWG); Single-Drive Mach Zehnder Modulator (SD-MZM); Data Amplifier (Data Amp); Erbium Doped Fiber Amplifier (EDFA); Variable Optical Attenuator (VOA); Real Time Oscilloscope (RTO).

## 5. Results and discussion

The obtained experimental results are shown in Fig. 5. Channel performance was determined using BER measurements (for the received optical power of ∼ 0 dBm), and results obtained are given in Fig. 5(a). For the case when optical carriers were modulated with uncoded PAM4, the performance of filtered modes fluctuated around the 7% FEC (BER=$3.8\times10^{-3}$[31]) limit, after transmission over 3 km of SSMF. Approximately less than a quarter of the channels exhibited performance better than 7% FEC. The variation in the BER performance of the different channels is matched with the variation of their measured RIN. Fig. 5(b) shows the measured average RIN (DC-5 GHz and DC-30 GHz) for selected modes. RIN measurements of individually filtered comb lines were carried out as described in [32,33] and the received optical power during the measurements was ∼ 1 dBm. The measured average RIN values for over 5 GHz bandwidth were in the range from -122 dB/Hz to -130 dB/Hz, whilst the average RIN was significantly lower when averaged over 30 GHz bandwidth, in the range from 136-139.5 dB/Hz. This indicates the presence of high MPN in this device, which is the limiting factor to enable satisfactory performance of multi-level amplitude modulation formats such as PAM4. The total measured average RIN (DC-30 GHz) for all channels was ∼ -143 dB/Hz. The measured average RIN (DC-30GHz) values are lower than the values specified in 40GBASE-LR4 and 100GBASE-LR4 standards, however it should be noted that these standards are given for lasers which have essentially constant RIN as a function of frequency and not the high level of low frequency RIN associated with the filtered lines from the Q-Dash mode locked laser. Degradation in performance can be observed in Fig. 5(a) for lower wavelength modes (corresponding to high mode number), due to the reduction of the OCNR as result of lower optical powers in the lower wavelength comb lines, and an associated increase of the RIN. Thus the last few modes exhibit performance around BER=$1\times10^{-2}$ for standard PAM4 modulation. Performance difference between back-to-back case and transmission after 3 km of SSMF was measured and determined to be negligible. Considering dispersion properties of SSMF, a minimum length of 5 km of SSMF is required to cause the penalty of 1-dB, in the case that there is no chromatic dispersion compensation or equalization in the system [34].

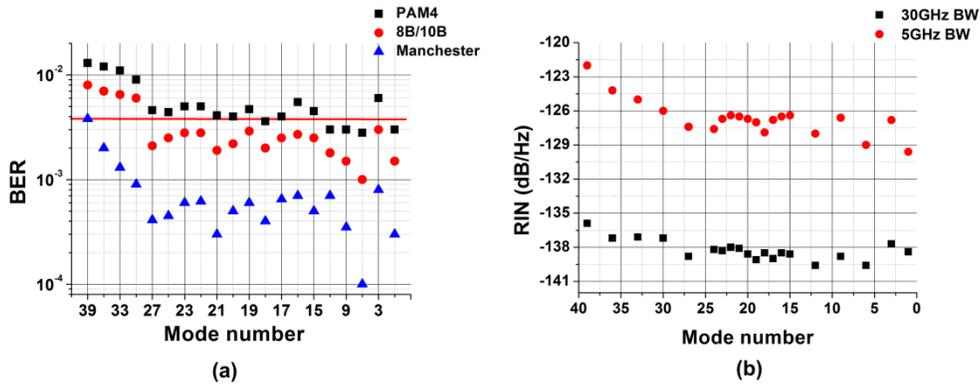

Fig. 5. a) Measured BER for selected channels after transmission over 3 km (b) the measured average RIN over 5 GHz and 30 GHz bandwidth for selected modes.

However, when 8B/10B encoding was applied, all modes (except the last few low wavelength modes) exhibited performance better than 7% FEC limit. FEC can be implemented before or after the 8B/10B encoder. Typically, FEC is implemented after the 8B/10B encoder where the data is scrambled to ensure that DC balancing is preserved [35]. However, implementation where FEC is employed before 8B/10B automatically ensures DC balancing and also guarantees the specified bandwidth of the spectral notch [36]. The spectral notch of a few hundred MHz, as shown in Fig. 2, enabled low frequency RIN mitigation and performance improvement compared to the case when uncoded PAM4 was utilized.

Further mitigation of MPN and performance improvement can be enabled if wider spectral notches are generated. This was possible when Manchester coding was applied, as shown in Fig. 2 and Fig. 5(a). Manchester encoded PAM4 enabled an order of magnitude performance improvement compared to regular PAM4. Moreover, all modes exhibited performance better than 7% FEC limit, at the expense of a halved net spectral efficiency for a given data rate due to the transitions overhead. Similar to the case for uncoded PAM4, the performance fluctuation of 8B/10B and Manchester PAM4 have a good match with the fluctuation of the measured RIN. Manchester encoding requires FEC encoder to be implemented before Manchester encoding block [37]. An ECL was used to set a benchmark performance of the system, as shown in Fig. 6. This figure shows the performance in terms of BER versus received optical power for modes 1 and 21, for uncoded and encoded PAM4. A received power of -7 dBm was required to achieve BER=$5\times10^{-6}$ for PAM4, and -8 dBm for 8B/10B and Manchester encoded PAM4 signal. A slight difference in performance at low received powers between the 8B/10B and Manchester encoded signals and the uncoded case is due to the superior suppression of the low frequency content that 8B/10B and Manchester encoding achieves. The low frequency suppression achieved by the 8B/10B and Manchester encoders can provide a better resilience towards signal distortions imposed by in-line equipment with minimum cut-off frequencies in the kHz to MHZ range [25]. The average RIN (DC-30 GHz) for the ECL was measured to be ~ -145 dB/Hz, with flat spectrum at low frequency range, as shown in Fig. 3(c). Furthermore, OCNR of the ECL was in excess of 65 dB which was measured with 0.02 nm RBW OSA. Consequently, the penalty in performance for Q-Dash carriers compared to the ECL, as shown in Fig. 6, is caused by the low frequency RIN and reduced OCNR of Q-Dash MLL. Furthermore, as it can be seen in Fig. 6, Manchester PAM4 exhibit the same performance at 7% FEC limit for ECL and Q-Dash devices, whilst 8B/10B PAM4 have ~ 2dB power penalty at 7% FEC for the Q-Dash device as compared to the ECL.

Considering that ~30 channels exhibited performance better than 7% FEC when 8B/10B

encoding was applied and ~40 channels when Manchester encoding was applied the potential aggregate raw capacities achievable after performing Nyquist shaping with the 33.6 GHz Q-Dash PMLL would be around 1.5-2 Tb/s. By using a Q-Dash PMLL with higher FSR, which would have less optical carriers compared to 33.6 GHz PMLL [4], 400 Gb/s and 1 Tb/s aggregate raw capacities would be achievable.

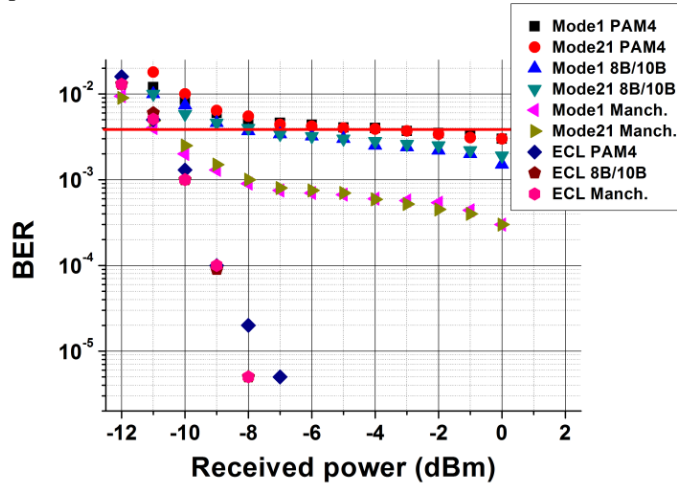

Fig. 6. BER versus received power for Mode 1 and Mode 21 for various scenarios, and ECL performance for 28 GBaud PAM4, 8B/10B PAM4 and Manchester PAM4.

## 6. Conclusions

In this paper we investigated low frequency RIN mitigation by employing 8B/10B and Manchester encoding with PAM4 modulation format. The suppression of low frequency spectral content has been used for the mitigation of MPN, by reducing the overlap between the modulation spectral content and the low-frequency RIN. To the best of our knowledge, this is the first demonstration of 8B/10B and Manchester encoded PAM4 signaling in optical communication systems. The RIN measurements of Q-Dash PMLL modes show high levels of RIN from DC to 4 GHz, but low levels of RIN for higher frequencies. The obtained performance for standard 28 GBaud PAM4 shows BER fluctuation around 7% FEC limit. 8B/10B encoding enabled performance improvement, and majority of channels exhibited performance better than 7% FEC limit, whilst Manchester encoding provided an order of magnitude improvement in the performance compared to the uncoded PAM4. Nevertheless, due to the coding overhead, net spectral efficiency of 8B/10B encoded PAM4 signal is reduced for 20%, and net spectral efficiency of Manchester encoded PAM4 signal is halved for a given data rate compared to uncoded PAM4. Finally, the effect of RIN on the system performance was demonstrated by examining the system performance when an ECL was used instead, and comparing to the results obtained with Q-Dash PMLL.

## Funding

This work was supported in part by the EU FP7 Project "Big Pipes", research grants from Science Foundation Ireland (SFI) and is co-funded under the European Regional Development Fund under Grant Numbers 13/RC/2077, 15/US-C2C/I3132, and 12/RC/2276.

## Acknowledgments

The authors would like to acknowledge Tektronix, Inc. for the use of their DPO77002SX real time oscilloscope.